\documentclass[12pt]{article}
\usepackage[paperwidth=8.5in,paperheight=11in,hmargin=1in,vmargin={1in,0.5in},
 footskip=0.5in,footnotesep=0.3in,nohead,dvips]{geometry}
\usepackage{graphicx}
\usepackage{amssymb,latexsym,amsthm,amsmath}
\usepackage{bm}
\newtheorem{prop}{Proposition}[section]
\newtheorem{dfn}[prop]{Definition}
\newtheorem{exm}[prop]{Example}
\newtheorem{thm}[prop]{Theorem}
\newtheorem{rmk}[prop]{Remark}

\newcommand{\inv}{a_0}

\begin{document}
\begin{center}\Large Invariant classification of the rotationally symmetric $R$-separable webs for the Laplace equation \\in Euclidean space \end{center}

\begin{center}\large Mark Chanachowicz\footnote{Department of Physics,
  University of Waterloo, Waterloo, Ontario, N2L~3G1, Canada, email:\ mchanach@math.uwaterloo.ca},
  Claudia M.\ Chanu\footnote{Dipartimento di Matematica, Universit\`a di Torino,
  via Carlo Alberto 10, 10123 Torino, Italia, email:\ claudiamaria.chanu@unito.it}, \\
  Raymond G.\ McLenaghan\footnote{Department of Applied Mathematics, University of
  Waterloo, Waterloo, Ontario N2L~3G1, Canada, email:\ rgmclena@uwaterloo.ca} \\
  October 24, 2007 \end{center}
\date{\today}  

\begin{quote}{\small\textbf{Abstract.}
An invariant characterization of the rotationally symmetric $R$-separable webs for the Laplace equation in 
Euclidean space is given in terms of invariants and covariants of a real binary quartic canonically associated to the characteristic conformal Killing tensor which defines the webs.} \end{quote}

\section{Introduction}
\setcounter{equation}{0}
The theme of this paper is the study of $R$-\textit{separability} for the Laplace equation
\begin{equation}\label{Leq}
\Delta{\psi}=\dfrac{\partial^2 \psi}{\partial x_1^2}+\dfrac{\partial^2 \psi}{\partial x_2^2}+\dfrac{\partial^2 \psi}{\partial x_3^2} =0,
\end{equation}
in three-dimensional Euclidean space $\mathbb E^3$, that is the search for curvilinear coordinates $(q^i)$ such that the equation can be split into a system of separated ordinary differential equations by assuming a solution of the form 
$$
\psi= R(q^1,q^2,q^3)\textstyle\prod_i \phi_i(q^i,c_a),
$$
where $R$ is some nowhere vanishing function on $\mathbb E^3$ and $\phi_i$ are functions of the single coordinate $q^i$ and a suitable set of constants $(c_a)$.  
B\^{o}cher \cite{Bocher} (see also \cite{Moon,Miller}) shows that the
equation separates in seventeen types of cyclidic coordinate systems. These systems are classified by group
theoretic methods by Boyer, Kalnins and Miller \cite{Boyer}. They explicitly described the \textit{equivalence problem} for the $R$-separable coordinate systems in terms of an orbit analysis of the algebra of the second order symmetry
operators of (\ref{Leq}) under the action of the conformal group of $\mathbb{E}^3$. Eleven of the coordinate systems
are \textit{simply separable} in the sense that they allow multiplicative separability in the ordinary sense (i.e., with $R=1$) of both (\ref{Leq}) and the Helmholtz equation
\begin{equation}\label{Heq}
\Delta{\psi}+\omega^2\psi=0,
\end{equation}
while the remaining six coordinate systems afford $R$-separability of the La\-pla\-ce equation only for non-trivial functions $R$. Boyer et al. also solve the \textit{canonical forms problem} by giving representative elements expressed in canonical
Cartesian coordinates for the pairs of commuting or $R$-commuting symmetry operators from each of the seventeen 
orbit classes they delineate.

The purpose of the present paper is to consider the problem of the characterization of the $R$-separable coordinate systems for (\ref{Leq}) from a geometric point of view. We present an invariant classification scheme for the
rotationally symmetric $R$-separable coordinate systems in terms of invariants of valence two \textit{conformal Killing
tensors} (CKT). These tensors are the geometric quantities in terms of which the second order symmetry
operators of (\ref{Leq}) are defined. Such a tensor with point-wise distinct eigenvalues and normal eigenvectors,
called a \textit{characteristic conformal Killing tensor},
defines a geometric structure called an \textit{$R$-separable web} which is a set of three mutually orthogonal foliations of the space by two-dimensional surfaces. Each parameterization of an $R$-separable web defines an $R$-separable coordinate system the coordinate surfaces of which
coincide with the leaves of the foliations. In this perspective the
characteristic CKTs play a central role in the theory of $R$-separation of variables. 

Our approach is an extension of that employed by Horwood, McLenaghan and Smirnov \cite{CMP} who give an invariant
classification of the eleven separable webs for the Hamilton-Jacobi equation for the geodesics and the Helmholtz equation
in $\mathbb{E}^3$ in terms of the invariants and reduced invariants of valence two Killing tensors.

The paper is organized as follows. In Sect. 2 we outline the theory of conformal Killing tensors defined on
pseudo-Riemannian spaces and give an overview of invariant theory for vector spaces of CKTs defined on 
$n$-dimensional flat spaces. Subsequently, we specialize the theory to the vector space of valence two trace-free CKTs defined on Euclidean space. In Sect. 3 we describe how the $R$-separable webs for the Schr\"{o}dinger equation may be characterized by CKTs and define the concept of web symmetry in terms of any characteristic CKT 
which defines the web. In Sect. 4 we derive the general form of the rotationally symmetric CKT in Euclidean space in addition to giving the characteristic CKTs
for each of the rotationally symmetric $R$-separable webs listed in \cite{Moon}. In Sect. 5 we give the group action, invariants and canonical forms for the
rotationally symmetric CKTs. We also show that the group action is equivalent to the classical action of 
$GL(2,\mathbb R)$
on real binary quartics. In Sect. 6 we give an invariant classification of the rotationally symmetric $R$-separable
webs in terms of invariants and covariants of the real binary quartic corresponding to a rotationally symmetric
CKT. In Sect. 7 we show that there exist no symmetric $R$-separable webs in Euclidean space other than the rotationally symmetric ones characterized in Sect. 6 or those that are conformally equivalent to simply separable symmetric webs. Sect. 8 contains the Conclusion.
 
\section{Invariant theory of conformal \\ Killing tensors}
\setcounter{equation}{0}
\subsection{Theory of conformal Killing tensors}
Let $(M,\mathbf{g})$ be a pseudo-Riemannian manifold with metric tensor $\mathbf{g}$. 
\begin{dfn} 
A {conformal Killing tensor of valence $p$} defined in $(M,\mathbf{g})$ is a symmetric $(p,0)$ tensor field $\mathbf{K}$ which satisfies the conformal Killing tensor equation
\begin{eqnarray}\label{CKT}
[\mathbf{g}, \mathbf{K}]=\mathbf{k}\odot\mathbf{g},
\end{eqnarray}
where $[,]$ denotes the Schouten bracket, $\mathbf{k}$ some symmetric tensor of type $(p-1,0)$, and $\odot$ denotes
the symmetric tensor product.  
\end{dfn}
The tensor $\mathbf{k}$ can be determined by contracting (\ref{CKT}) with the covariant metric tensor.

When $p=1$, $\mathbf{K}$ is said to be a \textit{conformal Killing vector} and (\ref{CKT}) reads
\begin{eqnarray}\label{CKVgen}
\mathcal{L}_{\mathbf{K}}\mathbf{g}=f\mathbf{g},
\end{eqnarray}
where $\mathcal{L}$ denotes the Lie derivative operator. With respect to a local system of coordinates $x^{i}$
(\ref{CKT}) may be written
\begin{eqnarray}\label{CKT2}
 \nabla_{(i_{1}}K_{i_{2}\ldots i_{p+1})}=k_{(i_{1}\ldots i_{p-1}}g_{i_{p}i_{p+1})},
\end{eqnarray}
where $\nabla$ denotes the covariant derivative with respect to the Levi-Cevita connection of $\mathbf{g}$. If $\mathbf{k} = 0$, in (\ref{CKT}), then $\mathbf{K}$ is said to be a \textit{Killing tensor}.

It follows from the properties of the
Schouten bracket that the set $CK^{p}(M)$ of all conformal Killing tensors of type $(p,0)$ forms a generally infinite dimensional 
vector space. However, it's important to note that
\begin{eqnarray}\label{CKT3}
\mathbf{K}' = \mathbf{K} + \mathbf{l}\odot\mathbf{g},
\end{eqnarray}
where $\mathbf{l}$ is any symmetric tensor of type $(p-2,0)$, also defines a CKT. This property may be used to define
the following equivalence relation on $CK^{p}(M)$:
\begin{eqnarray}\label{eqiv}
\mathbf{K}' \sim \mathbf{K} \Leftrightarrow \mathbf{K}' = \mathbf{K} + \mathbf{l}\odot\mathbf{g},
\end{eqnarray}  
Let $C\hat{K}^{p}(M)$ denote the set of equivalence classes of $CK^{p}(M)$. We may equip $C\hat{K}^{p}(M)$ with the structure of a vector space over the reals as follows. Let $\hat{\mathbf{K}}_{1}$ and $\hat{\mathbf{K}}_{2}$ $\in$ $C\hat{K}^{p}(M)$. Let $\mathbf{K}_{1}$ and $\mathbf{K}_{2}$ be representative elements of $\hat{\mathbf{K}}_{1}$ and $\hat{\mathbf{K}}_{2}$ respectively. Then $\hat{\mathbf{K}}_{1} + \hat{\mathbf{K}}_{2}$ is defined to be the equivalence class represented by ${\mathbf{K}}_{1} + {\mathbf{K}}_{2}$. Let $\mathbf{K}$ be representative of $\hat{\mathbf{K}}$ and 
$a \in \mathbb{R}$. Then $a\hat{\mathbf{K}}$ is defined to be the equivalence class represented by $a{\mathbf{K}}$. It is easy to check that these operations are well defined. Let $TCK^{p}(M)$ denote the vector space of trace-free conformal Killing tensors of type $(p,0)$. It is easily verified that $TCK^{p}(M)$ is canonically isomorphic to $C\hat{K}^{p}(M)$. A necessary and sufficient condition for an 
element of $C\hat{K}^{p}(M)$ to be represented by a Killing tensor is that there exists a type $(p - 2, 0)$ tensor $\mathbf{l}$ such that 
\begin{eqnarray}\label{KTcond}
[\mathbf{l}, \mathbf{g}] = \mathbf{k}.
\end{eqnarray}
For $p = 2$ the above equation may be written as 
\begin{eqnarray}\label{KTcond_2}
d\mathbf{l} = -\mathbf{k}.
\end{eqnarray}
The integrability condition for this equation is 
\begin{eqnarray}\label{KTcond_3}
d\mathbf{k} = 0.
\end{eqnarray}
By solving (\ref{CKT2}) for $\mathbf{k}$ we may write the integrability condition in component form as 
\begin{eqnarray}\label{KTcond_4}
K_{k[i;}{}^{k}{}_{j]} = 0.
\end{eqnarray}
This is a necessary and sufficient condition for $\hat{\mathbf{K}}$ to be represented by a Killing tensor.

We now study the behavior of the conformal Killing tensor $\mathbf{K}$ under a conformal transformation of the metric, which may be written as 
\begin{eqnarray}\label{conf_trans}
\tilde{\mathbf{g}} = e^{-2\sigma}\mathbf{g}.
\end{eqnarray}
By an easy calculation we find that
\begin{eqnarray}\label{conf_trans_2}
[\tilde{\mathbf{g}}, \mathbf{K}] = \mathbf{k}'\odot{\tilde{\mathbf{g}}},
\end{eqnarray}
where 
\begin{eqnarray}\label{conf_trans_3}
\mathbf{k}' = e^{-2\sigma}(\mathbf{k} - 2[\sigma,\mathbf{K}]).
\end{eqnarray}
This result shows that $\mathbf{K}$ is also a conformal Killing tensor for the conformally related metric $\tilde{\mathbf{g}}$. A necessary and sufficient condition that $\mathbf{K}$ is a Killing tensor with respect to $\tilde{\mathbf{g}}$ is that 
\begin{eqnarray}\label{conf_trans_4}
[\sigma,\mathbf{K}] = \frac{1}{2}\mathbf{k}.
\end{eqnarray}

\subsection{Conformal Killing tensors in spaces of zero curvature}
We now assume that the Riemann curvature tensor of $\mathbf{g}$ vanishes. In this case it has been shown by Eastwood \cite{Eastwood} that $C\hat{K}^{p}(M)$ is finite dimensional and that its dimension $d$ is given by 
\begin{eqnarray}\label{dimCKT}
d = \frac{(n + p - 3)!(n + p - 2)!(n + 2p - 2)(n + 2p - 1)(n + 2p)}{p!(p + 1)!(n-2)!n!},
\end{eqnarray}
for $n \geq 3, p \geq 1$.
Thus the general element of $C\hat{K}^{p}(M)$ is represented by $d$ arbitrary parameters $a^{1},\ldots,a^{d}$, with respect to an appropriate basis. 

Each element $h$ of the conformal group $CI(M)$ induces, by the push forward map, a non-singular linear transformation $\rho(h)$ of $C\hat{K}^{p}(M)$. It is implicit in the work of \cite{Eastwood} that the map 
\begin{eqnarray}\label{repn}
\rho:CI(M) \rightarrow GL(C\hat{K}^{p}(M))
\end{eqnarray}
defines a representation of $CI(M)$. Once the form of the general element $\hat{\mathbf{K}}$ of $C\hat{K}^{p}(M)$ is available with respect to some convenient coordinate system on $M$, the explicit form of the transformation $\rho(h)\hat{\mathbf{K}}$ (written more succinctly as $h \cdot \hat{\mathbf{K}}$) may be written explicitly in terms of the parameters $a^{1},\ldots,a^{d}$. We shall be particularly concerned with the smooth real-valued functions on $C\hat{K}^{p}(M)$ that are invariant under the group $CI(M)$. The precise definition of such $CI(M)$-invariant functions of $C\hat{K}^{p}(M)$ is as follows.
\begin{dfn} 
Let $(M,\mathbf{g})$ be a pseudo-Riemannian manifold with zero curvature. Let $p \geq 1$ be fixed. A smooth function $F: C\hat{K}^{p}(M) \rightarrow \mathbb R$ is said to be  a {$CI(M)$-invariant} of $C\hat{K}^{p}(M)$ if it satisfies the condition
\begin{eqnarray}\label{inv_def}
F(h\cdot\hat{\mathbf{K}}) = F(\hat{\mathbf{K}}),
\end{eqnarray}
for all $\hat{\mathbf{K}} \in C\hat{K}^{p}(M)$ and for all $h \in CI(M)$. 
\end{dfn}
The main problem of invariant theory is to describe the whole space of invariants of a vector space under the action of the group. To achieve this one has to to determine the set of \textit{fundamental invariants} with the property that any other invariant is an analytic function of the fundamental invariants (see \cite{Olver_2}). The fundamental theorem of invariants for a regular Lie group action \cite{Olver_2} determines the number of fundamental invariants needed to define the whole of the space of $CI(M)$-invariants.

\begin{thm}\label{fund_theorem}
Let $G$ be a Lie group acting regularly on an n-dimensional manifold $M$ with $s$-dimensional orbits. Then, in a neighborhood $N$ of each point $p \in M$, there exists $n - s$ functionally independent $G$-invariants\\
$\Delta_{1}, \ldots, \Delta_{n-s}$. Any other $G$-invariant $I$ defined near $p$ can be locally uniquely expressed as an analytic function of the fundamental invariants namely $I$ = $F(\Delta_{1}, \ldots, \Delta_{n-s})$.
\end{thm}

One of the standard methods for determining the invariants of $C\hat{K}^{p}(M)$ is to use the fact that the invariants of a function under an entire Lie group is equivalent to the invariants of the function under the infinitesimal transformation of the group given by the corresponding Lie algebra. The precise result is as follows \cite{Olver}

\begin{prop}
Let $G$ be a connected Lie group of transformations acting regularly on a manifold $M$. A smooth real valued function $F: M \rightarrow \mathbb R$ is $G$-invariant if and only if
\begin{eqnarray}\label{inv_prop}
\mathbf{v}(F) = 0,
\end{eqnarray}
for all $p \in M$ and for every infinitesimal generator $\mathbf v$ of $G$.
\end{prop}

In our application $G$ is 
the representation $\rho(CI(M))$ defined by (\ref{repn}) where the condition (\ref{inv_prop}) reads 
\begin{eqnarray}\label{gener}
\mathbf{U}_{i}(F) = 0, \ \ \ \ i = 1, \ldots, r,
\end{eqnarray}
where the $\mathbf{U}_i$ are vector fields which form a basis of the Lie algebra of the representation and $r = \mathrm{dim} \;  CI(M) = \frac{1}{2}(n + 1)(n + 2)$. This Lie algebra is isomorphic to the Lie algebra of $CI(M)$. Such a basis may be computed directly as the basis of the tangent space to $\rho(CI(M)$ at the identity if an explicit form of the representation is available. According to Theorem \ref{fund_theorem} the general solution of the system of first-order PDEs (\ref{gener}) is an analytic function $F$ of a set of fundamental $CI(M)$-invariants. The number of fundamental invariants is $d - s$, where $d$ is given by (\ref{dimCKT}) and $s$ is the dimension of the orbits of $\rho(CI(M))$ acting regularly on the space $C\hat{K}^{p}(M)$.

We are now ready to apply the above theory to the vector space $C\hat{K}^{2}(\mathbb{E}^{3})$. Recall the following well-known result from invariant theory \cite{Olver_2}.
\begin{thm}\label{orbits}
The orbits of a compact linear group acting in a real vector space are separated by the fundamental (polynomial) invariants.
\end{thm}

We first note that in our case the group is non-compact and so in order to distinguish between the orbits of $CI(\mathbb{E}^{3})$ acting on the vector space $C\hat{K}^{2}(\mathbb E^3)$ we need to employ a more elaborate analysis than a mere computation of a set of fundamental invariants.

\subsection{Construction of the general CKT in $\mathbb{E}^{3}$}  
We now specialize the general theory of the previous subsection to the vector space $C\hat{K}^{2}(\mathbb E^3)$ of conformal Killing tensors of type $(2,0)$ defined in Euclidean space $\mathbb{E}^{3}$. 

It is well-known \cite{Rani} that in $\mathbb{E}^{3}$, any CKT is expressible modulo a multiple of the metric as a sum of symmetrized products of conformal Killing vectors. A canonical basis of the Lie algebra of conformal Killing vectors in $\mathbb{E}^{3}$ with respect to a system of Cartesian coordinates $(x_i)$ may be written as 
\begin{eqnarray}\label{CKV}
&\mathbf{X}_i& = \frac{\partial}{\partial{x_i}},\nonumber\\
&\mathbf{R}_i& = \epsilon_{ijk}x_{j}\mathbf{X}_{k},\nonumber\\
&\mathbf{D}& = x_{i}\mathbf{X}_{i},\nonumber\\
&\mathbf{I}_i& = (2x_{i}x_{k} - \delta_{ik}x_{j}x_{j})\mathbf{X}_{k},
\end{eqnarray}
for $i = 1,2,3$, where $\epsilon_{ijk}$ is the Levi-Cevita tensor. We also note the commutation relations
\begin{eqnarray}\label{CKV_commut}
&[\mathbf{X}_{i},\mathbf{X}_{j}]& = 0,\nonumber\\
&[\mathbf{X}_{i},\mathbf{R}_{j}]& = -\epsilon_{ijk}\mathbf{X}_{k},\nonumber\\
&[\mathbf{R}_{i},\mathbf{R}_{j}]& = -\epsilon_{ijk}\mathbf{R}_{k},\nonumber\\
&[\mathbf{X}_{i},\mathbf{D}]& = \mathbf{X}_{i},\nonumber\\
&[\mathbf{R}_{i},\mathbf{D}]& = 0,\nonumber\\
&[\mathbf{I}_{i},\mathbf{I}_{j}]& = 0,\nonumber\\
&[\mathbf{X}_{i},\mathbf{I}_{j}]& = 2(\delta_{ij}\mathbf{D} - \epsilon_{ijk}\mathbf{R}_{k}),\nonumber\\
&[\mathbf{R}_{i},\mathbf{I}_{j}]& = -\epsilon_{ijk}\mathbf{I}_{k},\nonumber\\
&[\mathbf{D}, \mathbf{I}_{i}]& = \mathbf{I}_{i}.
\end{eqnarray}

We now determine the form of the general element of $TCK^{2}(\mathbb{E}^3)$. By (\ref{dimCKT}) $d = 35$. It is clear that a sum of symmetrized products of conformal Killing vectors is a conformal Killing tensor. It will be shown that all trace free conformal Killing tensors may be obtained in this way. We begin by writing
\begin{eqnarray}\label{general_symm_product}
\mathbf{K} &=& A_{ij}\mathbf{X}_{i}\odot{\mathbf{X}_{j}} + B_{ij}\mathbf{X}_{i}\odot{\mathbf{R}_{j}} + C_{ij}\mathbf{R}_{i}\odot{\mathbf{R}_{j}} + D_{i}\mathbf{X}_{i}\odot{\mathbf{D}} + E_{ij}\mathbf{X}_{i}\odot{\mathbf{I}_{j}}\nonumber\\                 &+&  F_{i}\mathbf{R}_{i}\odot{\mathbf{D}} + G_{ij}\mathbf{R}_{i}\odot{\mathbf{I}_{j}} + H\mathbf{D}\odot{\mathbf{D}} + L_{i}\mathbf{D}\odot{\mathbf{I}_{i}} + M_{ij}\mathbf{I}_{i}\odot{\mathbf{I}_{j}}.                  
\end{eqnarray}
The coefficients in (\ref{general_symm_product}) obey the following symmetry relations
\begin{eqnarray}
A_{ij} = A_{ji},\ \ \ \ C_{ij} = C_{ji},\ \ \ \ M_{ij} = M_{ji}.
\end{eqnarray}
Thus an upper bound for the dimension of $TCK^{2}(\mathbb{E}^3)$ is fifty five, which exceeds the required dimension. Indeed there exist the following six relations among the basis set of symmetric tensor products of conformal Killing vectors:
\begin{eqnarray}
&&{\mathbf{X}_{i}\odot{\mathbf{R}_{i}}} = 0,\nonumber\\
&&{\mathbf{I}_{i}\odot{\mathbf{R}_{i}}} = 0,\nonumber\\
&&\mathbf{D}\odot{\mathbf{D}} = {\mathbf{X}_{i}\odot{\mathbf{I}_{i}} + \mathbf{R}_{i}\odot{\mathbf{R}_{i}}},\nonumber\\
&&2\mathbf{R}_{i}\odot{\mathbf{D}} + \epsilon_{ikl}\mathbf{X}_{k}\odot{\mathbf{I}_{l}} = 0.
\end{eqnarray}
Consequently, the general element of $TCK^{2}(\mathbb{E}^3)$ may be written as
\begin{eqnarray}\label{special_symm_product}
\mathbf{K} &=& A_{ij}\mathbf{X}_{i}\odot{\mathbf{X}_{j}} + B_{ij}\mathbf{X}_{i}\odot{\mathbf{R}_{j}} + C_{ij}\mathbf{R}_{i}\odot{\mathbf{R}_{j}} + D_{i}\mathbf{X}_{i}\odot{\mathbf{D}}\nonumber\\
 &+& E_{ij}\mathbf{X}_{i}\odot{\mathbf{I}_{j}} + G_{ij}\mathbf{R}_{i}\odot{\mathbf{I}_{j}} + L_{i}\mathbf{D}\odot{\mathbf{I}_{i}} + M_{ij}\mathbf{I}_{i}\odot{\mathbf{I}_{j}},                 
\end{eqnarray}
where the coefficients $B_{ij}$ and $G_{ij}$ may be chosen to satisfy
\begin{eqnarray}
{B_{ii}} = 0,\nonumber\\
{G_{ii}} = 0.
\end{eqnarray}
In terms of the natural basis, ${\mathbf{X}_{i}\odot{\mathbf{X}_{j}}}$, the components of $\mathbf{K}$ are given by
\begin{eqnarray}\label{CKT_compts_1}
K_{ij} &=& A_{ij} + (B_{(i|k}\epsilon_{kl|j)} + D_{(i}\delta_{j)l})x_l \nonumber\\  &+& (C_{mn}\epsilon_{mk(i}\epsilon_{|nl|j)} + 2E_{(i|k|}\delta_{j)l} - E_{(ij)}\delta_{lk})x_{l}x_{k}\nonumber\\ &+& (2G_{mn}\epsilon_{mk(i}\delta_{j)l} - G_{m(i}\epsilon_{j)mk}\delta_{ln} 
+ 2L_{k}\delta_{in}\delta_{jl} - L_{(i}\delta_{j)n}\delta_{lk})x_{l}x_{k}x_{n}\nonumber\\ &+& (4M_{kl}\delta_{jn} - 4M_{k(i}\delta_{j)n}\delta_{li} + M_{ij}\delta_{kn}\delta_{li})x_{l}x_{k}x_{n}x_{i}. 
\end{eqnarray}
Next we impose the trace free condition namely that
\begin{eqnarray}
K_{ii} = 0.
\end{eqnarray}
This procedure yields the following additional fourteen relations between the coefficients of $\mathbf{K}$: 
\begin{eqnarray}\label{trace_free_cond}
A_{ii} &=& 0,\ \ \ D_{i} = B_{jk}\epsilon_{kji},\ \ \ E_{kk} = 2C_{kk},\nonumber\\
E_{(ij)} - \frac{1}{3}E_{kk}\delta_{ij} &=& \frac{1}{2}(C_{ij} - \frac{1}{3}C_{kk}\delta_{ij}),\ \ L_{i} = G_{lm}\epsilon_{mli},\ \ M_{ii} = 0.
\end{eqnarray}
We can use the above conditions to remove fourteen coefficients: to keep the expression of $K_{ij}$ as symmetric as possible, we chose to eliminate the $D_{i}$, $L_{i}$, and $C_{ij}$. Moreover, the matrices $A_{ij}$ and $M_{ij}$ are necessarily trace free (we recall that also $B_{ij}$ and $G_{ij}$ are chosen to be trace free).

In terms of the natural basis, ${\mathbf{X}_{i}\odot{\mathbf{X}_{j}}}$, the components of $\mathbf{K}$ are given by:
\begin{equation}\label{CKT_compts_2}
\begin{array}{l}
K_{ij} = A_{ij} + (B_{(i|k}\epsilon_{kl|j)} + B_{lm}\epsilon_{ml(i}\delta_{j)k})x_k \\[4pt]  
+ \;[(2E_{(mn)} - 1/2E_{pp}\delta_{mn})\epsilon_{mk(i}\epsilon_{|nl|j)} + 2E_{(i|k|}\delta_{j)l} - E_{(ij)}\delta_{lk}]x_{k}x_{l}\\ [4pt]
+\; [2G_{mn}\epsilon_{nk(i}\delta_{j)l} - G_{n(i}\epsilon_{j)nk}\delta_{lm} 
+\; 2G_{np}\epsilon_{pnk}\delta_{im}\delta_{jl} - G_{ab}\epsilon_{ba(i}\delta_{j)n}\delta_{lk}]x_{k}x_{l}x_{m}\\ [4pt] 
+\, (4M_{kl}\delta_{mn} - 4M_{k(i}\delta_{j)l}\delta_{mn} + M_{ij}\delta_{kl}\delta_{mn})x_{k}x_{l}x_{m}x_{n}. 
\end{array}
\end{equation}
Moreover, any CKT of $\mathbb{E}^3$ is equivalent to
$$
A_{ij}\mathbf{X}_{i}\odot{\mathbf{X}_{j}} + B_{ij}\mathbf{X}_{i}\odot{\mathbf{R}_{j}} + 
 + E_{ij}\mathbf{X}_{i}\odot{\mathbf{I}_{j}} + G_{ij}\mathbf{R}_{i}\odot{\mathbf{I}_{j}} + M_{ij}\mathbf{I}_{i}\odot{\mathbf{I}_{j}},
 $$
where $A_{ij}$, $M_{ij}$, $B_{ij}$ and $G_{ij}$ must be trace free matrices.

The condition (\ref{KTcond_4}) applied to (\ref{CKT_compts_2}) implies that
\begin{equation}
E_{[ij]}=0, G_{ij}=0, M_{ij}=0.
\end{equation}
It follows from the above that (\ref{CKT_compts_2}) reduces to 
\begin{equation}\label{KT_compts}
\begin{array}{l}
K_{ij} = A_{ij} + (B_{(i|k}\epsilon_{kl|j)} + B_{lm}\epsilon_{ml(i}\delta_{j)k})x_k \\[4pt]  
+ \;[(2E_{mn} - 1/2E_{pp}\delta_{mn})\epsilon_{mk(i}\epsilon_{|nl|j)} + 2E_{k(i}\delta_{j)l} - E_{ij}\delta_{lk}]x_{k}x_{l},\\ [4pt] 
\end{array}
\end{equation}
which is the trace free part of an ordinary Killing tensor.

\section{Applications of CKTs to the geometric theory of separation of variables}
It is well known that Killing tensors are deeply related with additive separation of variables for the Hamilton-Jacobi equation for the geodesics or a natural Hamiltonian in orthogonal coordinates (see \cite{Kalnins}, \cite{Benenti_Chanu} and references therein)
$$
g^{ii}(\partial_i W)^2+ V=E, \qquad E\in \mathbb R,
$$
as well as for multiplicative separation of the Schr\"odinger equation 
$$
\Delta\psi +(E-V)\psi=0, \qquad E\in \mathbb R,
$$
where $\Delta$ is the Laplace Beltrami operator.
Indeed, the existence of a coordinate system in which separation of variable occurs is equivalent (for $V=0$) to the existence of a Killing tensor $\mathbf K$ with real simple eigenvalues and normal eigenvectors, called a {\em characteristic Killing tensor}. The separable coordinate hypersurfaces are defined to be orthogonal to the eigenvectors of $\mathbf K$ (the existence of these surfaces is equivalent to the normality of the eigenvectors). The set of the coordinate hypersurfaces is called an {\em orthogonally separable web}. Any parametrization of it locally defines {\em orthogonally separable coordinates}.
Moreover, if $V\neq 0$, the potential $V$ must satisfy an additional compatibility condition 
also expressed in terms of the characteristic KT (\cite{Benenti}):
\begin{equation}\label{dKdV}
d(\mathbf K dV)=0,
\end{equation}
where $\mathbf K$ is interpreted as a linear operator on one-forms.
Finally, for the multiplicative separation of the Schr\"odinger equation the so called \textit{Robertson condition} must also hold: the Ricci tensor is diagonalised in the separable coordinates (\cite{Eisenhart}). Geometrically, this means that $\mathbf K$ and the Ricci tensor have the same eigenvectors (\cite{Benenti_Chanu}).  
The condition that the eigenvalues are real is automatically satisfied for positive definite metrics; recently, KT's with complex conjugate eigenvalues have also been used to separate variables for a natural Hamilton-Jacobi equation \cite{Giovanni_Luca}.  

Similar results also hold for conformal Killing tensors. 

\begin{dfn} 
A {characteristic CKT} is a valence two conformal Killing with real and simple eigenvalues and normal eigenvectors.
\end{dfn}

\begin{rmk} \rm
Any CKT equivalent to a characteristic one is characteristic. Hence, 
it is always possible to choose a characteristic CKT which is trace free.
\end{rmk}

The following result holds (see \cite{HJE}):

\begin{prop}
There exists an orthogonal coordinate system in which additive separation for the null geodesic Hamilton-Jacobi
equation, 
$$
g^{ii} (\partial_i W)^2=0,
$$
occurs, if and only if there exists a characteristic CKT $\mathbf K$ on $M$. The coordinates hypersurfaces are orthogonal to the eigenvectors of $\mathbf K$.
\end{prop}

\begin{dfn} 
We call a {conformally separable web} the set of hypersurfaces orthogonal to the eigenvectors of a characteristic CKT. Any coordinates associated with a conformally separable web are called 
{conformally separable coordinates.}
\end{dfn}

\begin{prop}
There exists an orthogonal coordinate system in which additive separation for the  
Hamilton-Jacobi equation with fixed value of the energy $E$,
$$
g^{ii} (\partial_iW)^2+V-E=0,
$$
occurs, if and only if there exists a characteristic CKT $\mathbf K$ on $M$ satisfying the compatibility condition
\begin{equation}\label{Vcompat}
d((E-V) \mathbf k^\flat+2\mathbf KdV)=0,
\end{equation}
where $\mathbf k^\flat$ is the one-form corresponding to the vector field $\mathbf k$ such that 
$[\mathbf{g}, \mathbf{K}]=\mathbf{k}\odot\mathbf{g}$ and $\mathbf K$ is considered as a linear operator on one-forms.
\end{prop}

\begin{proof}
According to \cite{HJE} we have that the natural Hamilton-Jacobi equation with fixed value $E$ of the energy is separable, if and only if there exist a function $f$ and a characteristic CKT $\mathbf K$ such that
$$
[\mathbf g, \mathbf K]=\frac{2}{E-V}(\mathbf K \nabla V + \nabla f)\odot \mathbf g, 
$$
where $\nabla$ is the gradient operator and $\mathbf K$ is considered as a linear operator on vector fields. 
By (\ref{CKT}), we have that $(E-V) \mathbf k -2 K\nabla V$ must be the gradient of a function and (\ref{Vcompat}) follows by passing from vector fields to one-forms.
\end{proof}

\begin{rmk} \rm
If $\mathbf K$ is a Killing tensor, the compatibility condition (\ref{Vcompat}) reduces to (\ref{dKdV})
\end{rmk}

In spite of the fact that the null geodesic equation is trivial for a positive definite metric, conformally separable coordinates are useful because they are the only ones in which a natural Hamiltonian with fixed value of the energy can be solved by additive separation of variables. Moreover, they are the only ones in which $R$-separation of the Laplace equation can occur. We recall the definition given in \cite{FERsep}:

\begin{dfn} 
We say that {multiplicative $R$-separation} of the Laplace equation 
$\Delta \psi=0$
occurs in a coordinate system $(q^i)$ if there exists a 
solution $\psi$ of the form 
\begin{equation}\label{Rdef}
\psi=R(q^1,\ldots, q^n)\prod_i \phi_i(q^i,c_a) \qquad (c_a)\in \mathbb R^{2n-1},
\end{equation}
satisfying the completeness condition
  $$
\mathrm{rank}\left[ \frac{\partial}{\partial c_a}\!\left(\frac{\phi_i^\prime}{\phi}\right)\ \frac{\partial}{\partial c_a}\!\left(\frac{\phi_i^{\prime\prime}}{\phi}\right)\right]=2n-1, 
\qquad 
a=1,\ldots, 2n-1, \quad i=1,\ldots, n.
$$ 
\end{dfn}

The following theorem holds \cite{FERsep, Kalnins}:
\begin{thm}
On a flat manifold, $R$-separation of the Laplace equation occurs in a coordinate system $(q^i)$ if and only if
the coordinates $(q^i)$ are orthogonal conformally separable coordinates.
The function $R$ is (up to separated factors) a solution of the first order system
$$
\partial_i \ln R= \frac 12 \Gamma_i,
$$
where $\Gamma_i=g^{hk}\Gamma_{hki}$ denotes the contraction of the Christoffel symbols with the metric.
\end{thm}

\begin{rmk} \rm
If the manifold is not flat, then conformal separability is a necessary, but no longer sufficient condition. To guarantee 
$R$-separation we also require the function
$\frac{\Delta R}R$ be of the form $g^{ii}f_i(q^i)$ for suitable functions of a single variable $f_i$.
\end{rmk}

\begin{dfn}
We call an {$R$-separable web} a conformally separable web if $R$-separation for the Laplace equation occurs in any associated coordinate system. 
\end{dfn}

In $\mathbb{E}^3$, every conformally separable web is an $R$-separable web for the Laplace equation.
This means that any characteristic CKT defines an $R$-separable web.
$R$-separable coordinates of $\mathbb{E}^3$ have been extensively studied by many authors (see B\^ocher\cite{Bocher}, Moon and Spencer \cite{Moon}, Boyer et al.\cite{Boyer}). The webs consist of families of confocal cyclides.

In Sect. 4 to 6 we restrict ourselves to the webs and associated characteristic CKTs admitting rotational symmetry.
To make the notion of a web-symmetry precise, we give the definition of invariance of conformal Killing tensors under one parameter groups of conformal transformations. 
\begin{dfn}
Let $\mathbf{K}$ denote a characteristic conformal Killing tensor on $(M,\textbf{g})$. Let $\phi_t$ denote a one parameter group of conformal transformations. The $R$-separable webs defined by $\mathbf{K}$ are said to be {$\phi_t$-symmetric} if and only if
\begin{eqnarray}\label{web-symm}
\phi_{t\ast}K = fK,
\end{eqnarray}
where $\phi_{t\ast}$ denotes the push-forward map and $f$ is some function.
\end{dfn}
The infinitesimal version of the above definition is given by the following proposition:
\begin{prop}\label{web-symm_2}
Let $\mathbf{V}$ be an infinitesimal generator of the one parameter group of conformal transformations $\phi_t$. Then $\phi_t$ is a web-symmetry of the $R$-separable web defined by a characteristic CKT $\mathbf{K}$ if and only if
\begin{eqnarray}\label{web-symm_3}
\mathcal{L}_{\mathbf{V}}\mathbf{K} = h\mathbf{K}
\end{eqnarray}
where $h$ is some function.
\end{prop}
If $\phi_t$ denotes a one-parameter group of homothetic transformations then the functions $f$ and $h$ are constant. Moreover, if $\phi_t$ denotes a one-parameter group of isometries then the functions $f$ and $h$ are identically zero.

\section{The set of rotationally symmetric CKTs.}\label{s_4}
\setcounter{equation}{0}
\subsection{Definition and construction of rotationally symmetric webs}
We apply (\ref{web-symm_3}) to compute the characteristic CKTs in $\mathbb{E}^3$ admitting a rotational symmetry around the $z$-axis, associated with rotationally $R$-separable webs.

\begin{prop}\label{p_rot}
A CKT $\mathbf K$ of $\mathbb{E}^{3}$ satisfies 
$$
\mathcal L_{ \mathbf R_3}\mathbf{K} = 0,
$$
and has normal eigenvectors if and only if it is equivalent to 
\begin{equation} \label{rotCKT}
M_{33}\mathbf I_3\odot\mathbf I_3 + L_3\mathbf D \odot \mathbf I_3 +  
H \mathbf D\odot \mathbf D +C_{33}\mathbf R_3\odot \mathbf R_3 + D_3 \mathbf D\odot \mathbf X_3 +A_{33}\mathbf X_3\odot \mathbf X_3.
\end{equation}
\end{prop}

\begin{proof}
The infinitesimal invariance condition is linear in the parameters and gives a nine dimensional linear subspace $L$ of $TCK^2(\mathbb{E}^{3})$.
To check the normality of the eigenvectors we apply the Tonolo-Schouten-Nijenhuis conditions (TSN-conditions)
\cite{Tonolo,Schouten,Nijenhuis} on the generic element of $L$.
The TSN conditions are both necessary and sufficient for a given symmetric tensor field to have normal eigenvectors. They read
\begin{eqnarray}
{N}^{l}_{[jk}g_{i]l} = 0, \nonumber\\
{N}^{l}_{[jk}K_{i]l} = 0, \nonumber\\
{N}^{l}_{[jk}K_{i]m}K^{m}_{l} = 0, 
\end{eqnarray}
where $N^{i}_{jk}$ are the components of the Nijenhuis tensor of $\mathbf K$ defined by
\begin{eqnarray*}
\textsl{N}^{i}_{jk} = K^{i}_{l}K^{l}_{[j,k]} + K^{l}_{[j}K^{i}_{k],l}.
\end{eqnarray*}
The TSN-conditions are verified in a six dimensional subspace $L'\subset L$ of $TCK^2(\mathbb{E}^{3})$.
However it appears that for the following calculations it is more effective to describe the elements of $L'$ as linear combinations of symmetric tensor products of CKVs that are not trace free CKTs.
\end{proof}
Let $RCK^2(\mathbb{E}^{3})$ be the subspace of $CK^2(\mathbb{E}^{3})$ of CKTs of the form (\ref{rotCKT}).
The free parameters describing a general element $\mathbf K\in RCK^2(\mathbb{E}^{3})$ are
\begin{equation}\label{par6}
(M_{33}, L_3, H, C_{33}, D_3, A_{33})
\end{equation}
and all the other forty nine coefficients of the general linear combination of symmetric products of CKVs 
(\ref{general_symm_product}) are null. 
Given any CKT in Cartesian coordinates satisfying $\mathcal L_{ \mathbf R_3}\mathbf K=0$, and the 
TSN-conditions, the value of the parameters (\ref{par6}) are determined as follows:
\begin{itemize}
\item
$M_{33}$ is 1/4 of the coefficient of $xyz^2$ in $K_{12}$;
\item
$L_3$ is 1/2 of the coefficient of $xyz$ in $K_{12};$ 
\item
$H$ is the coefficient of $xz$ in $K_{13}$; 
\item
$H-C_{33}$ is the coefficient of $xy$ in $K_{12}$; 
\item
$D_3$ is twice the coefficient of $x$ in $K_{13}$; 
\item
$A_{33}$ is the constant term of $K_{33}-K_{22}$. 
\end{itemize}
Since we are considering components (or functions of the components) which are not affected by the addition of
a multiple of the metric $f\mathbf g$, the six parameters are well defined, irrespective of whether one starts from a CKT in $TCK(\mathbb{E}^{3})$ or not.

\begin{rmk} \label{r_CVeigenv} \rm
Since $\mathbb{E}^{3}$ has dimension three, there is an equivalent way to characterize rotational $R$-separable webs.
Any rotational web contains a family of hypersurfaces made of half-planes issued from the rotation axis (the $z$-axis in our case).
These planes are orthogonal to the Killing vector $\mathbf R_3$. Hence $\mathbf R_3$ must be an eigenvector of the CKT defining the web. Moreover, this condition is also sufficient to ensure that the eigenvectors of $\mathbf K$ are normal. Indeed, one of them is the normal vector $\mathbf R_3$ and the other two are contained in the two-dimensional planes orthogonal to $\mathbf R_3$ and hence they are normal.
By imposing the condition
$$
(\mathbf{K}\cdot{\mathbf{R}_{3}})\times{\mathbf{R}_{3}} = 0,
$$
we find again the six dimensional linear subspace described by (\ref{rotCKT}).
\end{rmk}

Finally, in order to prove that the general rotational CKT (\ref{rotCKT}) is characteristic, we check that the eigenvalues are simple almost everywhere.
Since $\mathbf R_3$ is orthogonal to $\mathbf I_3$, $\mathbf D$, $\mathbf X_3$, we have
$$
\mathbf{K}\cdot{\mathbf{R}_{3}}= C_{33}(x^2+y^2)\mathbf R_3.
$$
Hence, $\mathbf R_3=\mathbf E_1$ is an eigenvector corresponding to the eigenvalue $\lambda_1=C_{33}(x^2+y^2)$.
The other two eigenvectors $\mathbf E_2$ and $\mathbf E_3$ are orthogonal to $\mathbf E_1$; they and
their corresponding eigenvalues do not depend on $C_{33}$.
Moreover, the associated eigenvalues are of the form
$$
\lambda_{2,3}=\frac{A\pm\sqrt{B}}2,
$$
where 
\begin{eqnarray}
 A =r^4M_{33}+zr^2L_3+r^2H+zD_3+A_{33}, \qquad( r^2=x^2+y^2+z^2 ) \qquad \\
 B =
(x^2+y^2)\left[r^2L_3+2zH+\dfrac{4z^2-r^2}{r^2}D_3+\dfrac{4z(2z^2-r^2)}{r^4}A_{33}\right]^2 + \qquad
\\
 \left[r^4M_{33}\!+\!zr^2L_3+(2z^2-r^2)H\!+\! \dfrac{z(4z^2-3r^2)}{r^2}D_3\!+\!\dfrac{r^4-8z^2(r^2-z^2)}{r^4}A_{33}\right]^2.
\nonumber
\end{eqnarray} 
Any change of the parameter $C_{33}$ does not affect the web; indeed, $\mathbf E_2$ and $\mathbf E_3$ do not involve $C_{33}$ (see also Sect. \ref{s:action}). Thus, it is always possible to choose $C_{33}$ such that $\lambda_1$ is different from $\lambda_2$ and $\lambda_3$ at any point outside of the $z$-axis. On the contrary for
$x=y=0$ we have
$$
\lambda_1=0,\qquad  \lambda_2=\tfrac 12(q(z)+|q(z)|),\qquad \lambda_3=\tfrac 12(q(z)-|q(z)|),
$$
with
\begin{equation}\label{poly_sing}
q(z)=M_{33}z^4 +L_3 z^3+H z^2+ D_3 z+ A_{33}.
\end{equation}
Thus, (at least) one of $\lambda_2$, $\lambda_3$ identically vanishes and all points of the $z$-axis are singular points of all rotational webs.
The singular points that are not on the rotation axis are those satisfying $\lambda_2=\lambda_3$, that is where $B=0$.

\begin{rmk} \label{r_sing}\rm 
The roots of (\ref{poly_sing}) are points on the $z$-axis where the three eigenvalues coincide and $\mathbf K$ is proportional to the metric tensor.  
The number of the roots $z_0$ of $q$ in $\mathbb{PR}^1$ (so that the point at infinity is also considered) and their multiplicity characterize the web from a geometric point of view.
\end{rmk} 
   
\begin{rmk} \rm
The knowledge of the eigenvalues of the characteristic tensor in a rotational web allows one to write the equations
of the (not planar) hypersurfaces (see \cite{Eigen})
The hypersurfaces $S_2$ orthogonal to $\mathbf E_2$ satisfy the equation
$$
\frac{\lambda_1-\lambda_3}{x^2+y^2}=h, \qquad h\in \mathbb R,
$$ 
while
the hypersurfaces $S_3$ orthogonal to $\mathbf E_3$ satisfy the equation
$$
\frac{\lambda_1-\lambda_2}{x^2+y^2}=h, \qquad h\in \mathbb R.
$$
It follows that the hypersurfaces have the form
$$
2(h-C_{33})(x^2+y^2)+A=\pm\sqrt{B},
$$
that is they are both described by the equation
\begin{equation}\label{sup23}
[2(h-C_{33})(x^2+y^2)+A]^2-B=0,
\end{equation}
but for different ranges of the value of $h$: we have surfaces of $S_2$ for $h<h_0$ and surfaces of $S_3$ for
$h>h_0$, respectively, where
$$
h_0=C_{33}-\frac{A}{2(x^2+y^2)}=C_{33}-\frac{r^4M_{33}+zr^2L_3+r^2H+zD_3+A_{33}}{x^2+y^2}.
$$
For $h=h_0$ we do not obtain a surface of the web because this value of the parameter $h$ would imply 
$B=0$, that is $\lambda_2=\lambda_3$.
Expanding the equations (\ref{sup23}) we arrive at
\begin{eqnarray}
& [4(H-C_{33}+h)M_{33}-L_3^2]r^4+[8M_{33}D_3-4(C_{33}-h)L_3]r^2z+
\nonumber \\ 
& [2L_3D_3-4(C_{33}-h)H]r^2 
+16M_{33}A_{33}z^2+4(C_{33}-h)^2(x^2+y^2)+
\\
& [8L_3A_{33}-4(C_{33}-h)D_3]z-D_3^2+4(H-C_{33}+h)A_{33}=0, \nonumber
\end{eqnarray}
which represents two families of confocal cyclides, one for $h>h_{0}$ and one for $h<h_{0}$.
\end{rmk} 

\subsection{Characteristic CKTs of the known $R$-separable \\ rotational coordinate systems}
Table 1 contains the parameters of a characteristic CKT corresponding to each of the rotational $R$-separable coordinates listed in Moon and Spencer's book \cite{Moon}. 
We briefly describe how they are determined (for further details, such as plots, transformation laws to Cartesian coordinates, components of the metric tensor in these coordinates, separated equations etc.,\ see \cite{Moon} or \cite{Bocher}). 
The CKTs are constructed from the 
St\"ackel matrices that are associated with each system of coordinates in \cite{Moon}.

Recall that a {\em St\"ackel matrix} is a regular matrix of functions $S_{ij}$ depending on the single variable $q^i$ corresponding to the row index $i$ of the element. 
One row (the first in the examples in \cite{Moon}) of the inverse of the
St\"ackel matrix contains the components of the contravariant metric tensor in the $R$-separable coordinates, while the other two rows are made of the components of two CKTs with common eigenvectors orthogonal to the web hypersurfaces. Moreover, there is always a real linear combination of these two tensors which provides a characteristic tensor of the web (see \cite{HJE}).

\begin{table}
\begin{tabular}[b]{|l|c|c|c|c|c|c|}
\hline
Coordinates & $M_{33}$ & $L_3$ & $H$ & $C_{33}$ & $D_3$ & $A_{33}$ \vphantom{$\dfrac 12$}  \\ 
 \hline
 Bi-cyclide & $-\frac {k^2}{a^2}$ & 0 & $1+k^2$ & $1+k^2$ & 0& $-{a^2}$ \vphantom{$\dfrac 12$} \\ 
 \hline 
\parbox{1.8cm}{\vskip 3pt Flat-ring \\  cyclide  \vskip 3pt}& 
$\frac {k^2}{a^2}$ & 0 & $1+k^2$ & $0$ & 0& ${a^2}$  \vphantom{$\dfrac 12$} \\ \hline
 Disk cyclide & $-\frac {k^2}{a^2}$ & 0 & $1-2k^2$ & $0$ & 0& $\scriptstyle{a^2(1-k^2)}{}$ 
\vphantom{$\dfrac 12$} \\ \hline
 Cap cyclide  & $\frac{a^2(1+k)^2}{k}$ & 0 & $\frac{4k-(k-1)^2}2$ & $\frac{-(k-1)^2}{2}$ & 0 & $\frac {k(k+1)^2}{16a^2}$  
 \vphantom{$\dfrac 12$} \\  \hline
Toroidal & $\frac 1{4a^2}$ & 0 & $\frac 12$ & $\frac 12$ & 0& $\frac{a^2}4$  
\vphantom{$\dfrac 12$} \\  \hline
Bispherical & $-\frac 1{4a^2}$ & 0 & $\frac 12$ & $\frac 12$ & 0& $-\frac{a^2}4$ 
\vphantom{$\dfrac 12$} \\ \hline
\parbox{2cm}{\vskip 3pt Inverse \\  prolate\\ spheroidal \vskip 3pt}  & 
$\frac 1{a^2}$ & 0 & -1 & 0 & 0 & 0 \\ \hline
\parbox{2cm}{\vskip 3pt Inverse \\  oblate\\ spheroidal  \vskip 3pt} & 
$-\frac 1{a^2}$ & 0 & -1 & 0 & 0 & 0  \\ [8pt] \hline
\parbox{2cm}{\vskip 3pt Tangent\\ spheres  \vskip 3pt} & 1& 0 & 0 & 0 & 0 & 0 \\  \hline
Cardioid & 0 & 1 & 0 & 0 & 0 & 0   \vphantom{$\dfrac 12$} \\ \hline
\parbox{2cm}{\vskip 3pt Prolate\\ spheroidal  \vskip 3pt} & 0 & 0 & -1 & 0 & 0 & $a^2$   \\   \hline
\parbox{2cm}{\vskip 3ptOblate\\ spheroidal  \vskip 3pt}& 0 & 0 & 1 & 0 & 0 & $a^2$ \\  \hline
Spherical & 0 & 0 & 1 & -1 & 0 & 0 \vphantom{$\dfrac 12$} \\ \hline
Parabolical & 0 & 0 & 0 & 0 & 1 & 0 \vphantom{$\dfrac 12$} \\  \hline
Cylindrical & 0 & 0 & 0 & -1& 0 & 1 \vphantom{$\dfrac 12$} \\   \hline
\end{tabular}
\caption{Characteristic CKT of rotationally symmetric $R$-separable webs}
\end{table}

For each row of the inverse of the St\"ackel matrix we construct the conformal Killing tensors in the $R$-separable coordinates, then
the parameters (\ref{par6}) are determined by transforming the tensor to Cartesian coordinates and comparing with the Cartesian components of the general rotationally symmetric CKT (\ref{rotCKT}).
For all the coordinate systems considered in \cite{Moon} the tensor corresponding to the third row of the inverse St\"ackel matrix is 
$\mathbf{R}_3 \odot \mathbf{R}_3$.
In most of the examples, the other tensor is a characteristic tensor of the web so its parameters appear unchanged in the Table 1. 
On the contrary, the tensors arising from the St\"ackel matrices given in \cite{Moon} for Spherical, Tangent spheres and Cylindrical coordinates have $C_{33}=0$, so they are not characteristic CKTs. In order to get a characteristic CKT associated with these webs we add a suitable multiple
of the tensor $\mathbf{R}_3\odot\mathbf{R}_3$: that is, we change the value of $C_{33}$ in Table 1.

The first four coordinate systems have transformation laws to Cartesian coordinates involving Jacobi elliptic functions.  The parameter $a$ is a scaling parameter, while the parameter $k\in (0,1)$ is the parameter of the Jacobi elliptic functions.

\section{Group action preserving rotationally \\ symmetric CKTs} \label{s:action}

\subsection{The group and its one-parameter subgroups}\label{ss:5.1}
In order to classify the different types of $R$-separable webs admitting a rotational symmetry, we consider transformations acting on $CK^2(\mathbb{E}^3)$ which preserve the space $RCK^2(\mathbb{E}^3)$ of the rotationally symmetric CKTs defined in Sect.
\ref{s_4}. For this purpose, we use a group $G$ that is generated by five one-parameter transformations and a discrete transformation.
Three of the one-parameter transformations are induced on $RCK^2(\mathbb{E}^3)$ by conformal transformations
of $\mathbb{E}^3$ mapping the $z$-axis into itself. The other two are transformations of the CKT that do not change the corresponding web.

The five continuous transformations to be taken into account are 
\begin{enumerate}
\item
The change of the tensor under a continuous inversion along the $z$-axis parameterized by $a_{0}$:
$$\phi_0:(x,y,z)\to
\left(\frac x{1+2\inv z+\inv^2r^2}, \frac{y}{1+2 \inv z+\inv^2r^2}, \frac{z+\inv r^2}{1+2\inv z+\inv^2r^2}\right), 
$$ where $r^2=x^2+y^2+z^2$.
\item
The change of the tensor under a translation along the $z$-axis parameterized by $a_1$: $$\phi_1:(x,y,z)\to(x,y,z+a_1).$$
\item
The change of the tensor under a dilation of the space with singular point at the origin parameterized by $a_2$:  
$$\phi_2:(x,y,z)\to(a_2 x,a_2 y,a_2 z), \qquad (a_2 \neq 0).$$
\item
The multiplication of the tensor by a non-zero scalar $a_3$:
$$ \mathbf K \to a_3 \mathbf K, \qquad (a_3\neq 0).$$
\item
The addition to the tensor of a multiple of $\mathbf R_3 \odot \mathbf R_3$:
$$ \mathbf K \to \mathbf K + a_4 \mathbf R_3 \odot \mathbf R_3.$$
\end{enumerate}
Moreover, the discrete transformation considered, is the one induced by the inversion $I$ with respect to the 
unit sphere with centre at the origin
\begin{equation}\label{inversion}
I:(x,y,z)\to
\left(\frac x{x^2+y^2+z^2}, \frac{y}{x^2+y^2+z^2}, \frac{z}{x^2+y^2+z^2}\right). 
\end{equation} 
Note that $I^{-1}=I$ and that for the continous inversion $\phi_0$ we have $\phi_0=I^{-1}\circ \phi_1\circ I$, where $\phi_1$ is the transaltion along the $z$-axis.

\begin{rmk} \rm
The addition of the metric $\mathbf{g}$ and the transformation induced by the rotation around the $z$-axis are not relevant, since they do not modify the parameters (\ref{par6}) defining the tensor. 
\end{rmk}

\subsection{Group action, invariants and canonical forms}
Let $G$ be the group generated by the above described transformations. Since the discrete inversion is included, $G$ is not connected.  
Moreover, two of the continuous one-parameter transformations are defined only for values of the parameter in $\mathbb R -\{0\}$, so that the connected component of $G$ containing the identity is characterized by $a_2>0$ and $a_3>0$. Two other discrete transformations are implicitly included in $G$: the change of sign of the tensor (for $a_3=-1$) and the transformation induced by the symmetry around the origin in $\mathbb{E}^3$
(for $a_2=-1$).

The effect of the inversion around the unit sphere on the coefficients (\ref{par6}) of $\mathbf K \in RCK^2(\mathbb{E}^3)$ is given by
 \begin{equation} \label{act_disc_inv}
\begin{array}{l}
\tilde M_{33}= A_{33},\\
\tilde L_3=  D_3, \\
\tilde H= H,  \\
\tilde C_{33}=C_{33},  \\
\tilde D_3=  L_3,   \\
\tilde A_{33}= M_{33}. \\
\end{array}
 \end{equation}
The equations of the action generated by the five continuous transformations acting on (\ref{par6}) are
\begin{eqnarray*}
&\tilde M_{33}= a_3\dfrac{P(\inv)}{a_2^2},\\
&\tilde L_3=  a_3\dfrac{-4a_1P(\inv)-a_2P^{(1)}(\inv) }{a_2^2},  \\
&\tilde H= a_3\dfrac{6a_1^2P(\inv)+3a_1a_2P^{(1)}(\inv)+a_2^2P^{(2)}(\inv)}{a_2^2},  \\
&\tilde C_{33}=a_4 + a_3C_{33}+a_3\dfrac{6a_1^2P(\inv)+3a_1a_2P^{(1)}(\inv)+a_2^2(P^{(2)}(\inv)-H)}{3a_2^2},  \\
&\tilde D_3=  a_3\dfrac{ -4a_1^3P(\inv) -3a_1^2a_2P^{(1)}(\inv)-2a_1a_2^2P^{(2)}(\inv) -a_2^3 P^{(3)}(\inv)}{a_2^2},   \\
&\tilde A_{33} = a_3\dfrac{a_1^4P(\inv) +a_1^3a_2P^{(1)}(\inv)+\ldots +a_1a_2^3 P^{(3)}(\inv)+a_2^4 P^{(4)}(\inv)}{a_2^2}, \\
\end{eqnarray*}
where 
\begin{equation}\label{poly_act}
P(\inv)= A_{33}\inv^4 -D_3\inv ^3 +H\inv^2 -L_3 \inv +M_{33},
\end{equation}
and
$$
P^{(n)}=\dfrac 1{n!}\dfrac{d\,^nP}{(d\inv )^n}.
$$ 
Since $C_{33}$ and $a_4$ are involved only with $\tilde C_{33}$, and $C_{33}$ is unchanged by the discrete inversion (\ref{act_disc_inv}),
we can disregard $C_{33}$ (which can be made equal to any fixed constant by choosing a particular value for $a_4$). Then
we consider the
reduced action on the vector subspace of $RCK^2(\mathbb E^3)$ defined by the five parameters
\begin{equation}\label{par5}
(M_{33}, L_3, H, D_3, A_{33})
\end{equation}
of the subgroup $G'$ of $G$ defined by $a_4=0$:
\begin{eqnarray}
&\tilde M_{33}= a_3\dfrac{P(\inv)}{a_2^2} \label{M33n}, \\
&\tilde L_3=  a_3\dfrac{-4a_1P(\inv)-a_2P^{(1)}(\inv) }{a_2^2},  \\
&\tilde  H= a_3\dfrac{6a_1^2P(\inv)+3a_1a_2P^{(1)}(\inv)+a_2^2P^{(2)}(\inv)}{a_2^2},  \\
&\tilde  D_3\!=\!  a_3\dfrac{ -4a_1^3P(\inv) -3a_1^2a_2P^{(1)}(\inv)-2a_1a_2^2P^{(2)}(\inv) -a_2^3 P^{(3)}(\inv)}{a_2^2},   \\
&\tilde A_{33}\!=\! a_3\dfrac{a_1^4P(\inv) +a_1^3a_2P^{(1)}(\inv)+\ldots +a_1a_2^3 P^{(3)}(\inv)+a_2^4 P^{(4)}(\inv)}{a_2^2}.   \label{A33n}
\end{eqnarray}
It appears that the building blocks of the action equation is the polynomial (\ref{poly_act}) and its derivatives.

\begin{rmk} \rm
If we denote the parameters (\ref{par5}) by $\alpha^i$ ($i=0,\ldots,4$), setting $\alpha^4=M_{33}$, $\alpha^3=L_3$, $\alpha^2=H$, $\alpha^1=D_3$, $\alpha^0=A_{33}$), then their transformation laws under the action can be written in a compact formal way as
$$
\tilde{\alpha}^{4-i}=\dfrac{a_3}{a_2^2}\;{\sum_{h=0}^i(-1)^i 
\left(\!\!\!\begin{array}{c}
{\scriptstyle{4-h}} \\ {\scriptstyle i}
\end{array}\!\!\!\right)
P^{(h)}(\inv)a_1^{i-h}a_2^h}, \qquad i=0,\ldots,4. 
$$
\end{rmk}

\begin{thm}
Let $G_1$ be the subgroup of $G'$ defined by $a_3>0$. Then, the action of $G_1$ on $(\ref{par5})$ given by $(\ref{M33n}--\ref{A33n})$ and $(\ref{act_disc_inv})$ is equivalent to the classical action of $GL(2,\mathbb R)$ on real binary quartics.
\end{thm}
\begin{proof}
Consider the following binary quartic constructed from the five coefficients (\ref{par5}) of the CKT: 
\begin{equation}\label{quart}
Q(X,Y)= M_{33} X^4+L_3X^3Y+HX^2Y^2 +D_3XY^3+A_{33}Y^4.
\end{equation} 
By inserting the linear transformation of the variables $(X,Y)$
$$
X= \alpha \bar {X} +\beta \bar{Y},\qquad Y=\gamma \bar{X} +\delta \bar{Y},
$$
with $(\alpha\delta-\beta\gamma) \neq 0$, in (\ref{quart}), 
we obtain a new quartic $\bar{Q}(\bar{X},\bar{Y})$ whose coefficients $\bar{M}_{33},\ldots, \bar{A}_{33}$ 
depend on the $GL(2,\mathbb R)$ matrix $$M=\left[\begin{matrix}\alpha & \beta \\ \gamma & \delta\end{matrix}\right]$$ and on the coefficients of $Q$ (${M}_{33},\ldots, {A}_{33}$). 
Since we assume $a_3>0$, by setting 
$$
\alpha=\sqrt[4]{a_3a_2^2}, \quad \beta=-a_1\sqrt[4]{a_3a_2^2}, \quad \gamma =-a_0\sqrt[4]{a_3a_2^2}, \quad  
\delta= (a_1a_0+a_2)\sqrt[4]{a_3a_2^2},
$$
we obtain equations $(\ref{M33n}--\ref{A33n})$. The regularity of $M$ follows from 
$(\alpha\delta-\beta\gamma)=\sqrt{a_3}a_2{|a_2|}\neq 0$, since $a_2{a_3} \neq 0$.
Furthermore, setting $\alpha=\gamma=0$, and $\beta=\delta=1$, we recover (\ref{act_disc_inv}).
Conversely, we prove that for any transformation of the quartic we can associate a transformation of $G_1$. We distinguish two cases: for $\alpha\neq0$, by setting
$$
a_0=-\gamma \alpha^{-1},\quad a_1=-\beta \alpha^{-1}, \quad a_2=(\alpha\delta-\beta\gamma) \alpha^{-1},
\quad a_3=(\alpha\delta-\beta\gamma)^2,
$$
into $(\ref{M33n}--\ref{A33n})$ we obtain the action of $M$ on the quartic form. The fact that $a_2{a_3}\neq 0$ follows from the regularity of the matrix $M$. If $\alpha=0$, we apply first the discrete inversion $(\ref{act_disc_inv})$
on the parameters of the CKT, that is we multiply $M$ by $\left[\begin{matrix} 0 & 1 \\ 1 & 0\end{matrix}\right]$ on the left. In this way we obtain a new matrix $M_1$ with $\alpha_1= \gamma \neq 0$ since $M$ is regular, and thus revert to the previous case.
\end{proof}

As an immediate consequence of the theorem we are able to determine the invariant of the action and the list of canonical forms which are given in the following propositions.

\begin{prop}
The only independent differential invariant of the action of $G$ on $RCK^2(\mathbb{E}^3)$ is
$$F=\frac{I^3}{J^2},$$ 
where the functions 
$$
I=12A_{33}M_{33}-3L_3D_3+H^2,
$$
$$
J=72A_{33}M_{33}H-27A_{33}L_3^2-27D_3^2M_{33}+9D_3L_3H-2H^3,
$$
are relative invariants of the action of $G$ and independent differential invariants for the action of the subgroup of $G$ defined by $a_3=1$.
\end{prop}
\begin{proof}
The functions $I$ and $J$ are the fundamental invariants (of weight 4 and 6 respectively) of the binary quartic form (\ref{quart}) \cite{Gurevich},\cite{Olver_2}.
\end{proof}

\begin{prop}
Each CKT of $RCK^2(\mathbb{E}^3)$ is equivalent under the action of $G$ to one of the following
representatives:
\begin{eqnarray}
I. &\mathbf{I}_3\odot \mathbf{I}_3+ \mu\mathbf{D}\odot \mathbf{D} +\mathbf{X}_3\odot \mathbf{X}_3, \qquad &\mu \in \mathbb R, \label{cycl_tor}
\\
II. &\mathbf{I}_3\odot \mathbf{I}_3+ \mu\mathbf{D}\odot \mathbf{D} -\mathbf{X}_3\odot \mathbf{X}_3, \qquad &\mu \in \mathbb R,\label{diskcycl}
\\
III. &\mathbf{I}_3\odot \mathbf{I}_3+\nu\mathbf{D}\odot \mathbf{D} , \qquad &\nu=\pm 1,
\\
IV. &\mathbf{D}\odot \mathbf{I}_3, &
\\
V. &\mathbf{I}_3\odot \mathbf{I}_3. &
\end{eqnarray}
\end{prop}

\begin{proof}
Starting from the list of canonical forms of real binary quartics (given for instance in \cite{Gurevich}), we combine those differing only by sign. We remark that for $\mu= 2$ the canonical form $I.$ 
is equivalent to $\mathbf{D}\odot\mathbf{D}$. 
\end{proof}

\begin{rmk} \rm
The action of $G$ over $RCK^2(\mathbb{E}^3)$ has infinitely many orbits. However, the
tensors in (\ref{cycl_tor}) and (\ref{diskcycl}) are not pairwise inequivalent for all values of $\mu$:
for $\mu\neq \pm 2$ there 
exists a finite number of $\mu'$ such that the corresponding tensors are pairwise equivalent (see \cite{Gurevich}). 
\end{rmk}

\section{Invariant classification of the $R$-separable rotationally symmetric webs}

The polynomial $P$ defined in (\ref{poly_act}) as the building block of the action equations
(\ref{M33n}--\ref{A33n})
is deeply related to the polynomial $q$ (\ref{poly_sing}). Indeed, we have
$P(X)=X^4 q(-1/X).$
Moreover $q$ is the inhomogeneous polynomial corresponding to the quartic binary form $Q$
(\ref{quart}). 

The roots of $q$ are the points on the $z$-axis where all the eigenvalues of $\mathbf K$ coincide (see Remark \ref{r_sing}).
The conformal transformations  $\phi_0$, $\phi_1$, $\phi_2$ and  $I$ described in Sect. \ref{ss:5.1}
map the $z$-axis to itself 
with a one to one correspondence (if we include also the point at infinity).
Thus two distinct points cannot be made coincident or removed. 
This provides the geometric interpretation of the fact that the invariants of $Q$ are invariants of the CKT defining the web. The meaning of $q$ in terms of invariant theory is made more precise in the following proposition.   

\begin{prop} 
The polynomial $q(z)=M_{33}z^4 +L_3 z^3+H z^2+ D_3 z+ A_{33}$ is a relative covariant of the induced extended action 
on $C\hat K^2(\mathbb{E}^3)\times \mathbb{E}^3$ restricted on the invariant subset $S_0=\{x=y=0\}$.
\end{prop} 

\begin{proof}
The equations of the extended action are (\ref{M33n}--\ref{A33n}) together with
\begin{eqnarray*}
&\tilde{x}=\dfrac{a_2x}{(\inv z+1)^2+\inv^2(x^2+y^2)}, \\
& \tilde{y}=\dfrac{a_2y}{(\inv z+1)^2+\inv^2(x^2+y^2)}, \\
& \tilde{z}=a_2\dfrac{z + \inv^2(x^2+y^2+z^2)}{(\inv z+1)^2 + \inv^2(x^2+y^2)}+a_1.
\end{eqnarray*}
The subset $S_0=\{x=y=0\}$ is an invariant subset of the extended action.
Moreover, on $S_0$ the transformation law for $z$ reduces to the linear fractional transformation
\begin{equation}\label{zn}
\tilde{z}=\frac{(a_2+a_1\inv) z+a_1}{\inv z+1} \qquad (a_2\neq 0),
\end{equation}
which is the general linear transformation on $\mathbb{RP}^1$ (see \cite{Olver_2}).
Let $\tilde{q}(\tilde{z})$ be the polynomial we obtain by inserting (\ref{M33n}--\ref{A33n}) and (\ref{zn}) in 
(\ref{poly_sing}). We obtain
\begin{equation}\label{covar}
(\inv z+1)^4\tilde{q}(\tilde{z})={a_3a_2^2}q(z),
\end{equation}
that is (up to $a_3$) a covariant of weight two of the action. 
For the discrete inversion (mapping $z$ into $\tilde{z}=1/z$), we immediately see that it maps $q(z)$ to $\tilde{q}(\tilde{z})=\frac{q(z)}{z^4}$
\end{proof}

Equation (\ref{covar}) shows that the number and multiplicity of the real roots of $q(z)$ (that is the
number and multiplicity of the real linear factors of $Q$) 
are invariant with respect to the group action. Hence they can be used to define and classify the different types of webs.

\begin{dfn} 
We say that two rotationally symmetric $R$-separable webs are of the same type if the   
polynomials associated with the corresponding characteristic CKT have the same number and multiplicity of real roots. 
\end{dfn}

Thus we have reduced the classification of rotational $R$-separable webs to the classical classification of real binary quartics (see \cite{Olver_2}, \cite{Gurevich}). 

We have nine types of webs, listed in Table 2.

\begin{table}
\begin{tabular}[c]{|l|l|l|}
\hline
 \parbox{2.9cm}{\vskip 6pt Associated web \vskip 6pt} &  roots of $q$  & canonical form of $\mathbf K$ \\ 
 \hline
\parbox{2.5cm}{ \vskip 4pt Bi-cyclide \vskip 3pt} & 4 real distinct roots & $I.$ for $\mu<-2$ \\ 
 \hline 
 Flat-ring  cyclide &  \parbox{3.5cm}{\vskip 2pt 4 distinct complex \\ conjugate roots \vskip 2pt} & $I.$ for $\mu>-2$, $\mu \neq 2$  \\   
 \hline
  Disk cyclide   & \parbox{3.5cm}{\vskip 2pt 4 distinct roots,  2 real, 2 complex conjugate} & $II.$  \\
  \hline
 \parbox{2.7cm}{\vskip 3pt Inverse prolate \\ spheroidal \vskip 3pt} &   \parbox{3.5cm}{\vskip 3pt 1 double real root, \\ 2 distinct real roots \vskip 3pt} & $III.$ for $\nu=-1$ \\  
 \hline
 \parbox{2.5cm}{\vskip 3pt Inverse oblate \\ spheroidal \vskip 3pt}  & 
  \parbox{3.5cm}{\vskip 3pt 1 double real root, \\ 2 distinct complex \\ conjugate roots \vskip 3pt}& $III.$ for $\nu=1$ \\
 \hline
 Toroidal &  \parbox{3.5cm}{\vskip 3pt 2 double complex \\ conjugate roots} & $I.$ for $\mu=2$\\   \hline
\parbox{2.5cm}{ \vskip 4pt Bispherical \vskip 3pt} & 2 double real roots & $I.$ for $\mu=-2$ \\  \hline
 Cardioid &  \parbox{3.5cm}{\vskip 3pt  1 triple (real) root \\ 1 simple real root \vskip 3pt}& $IV.$ \\
 \hline
\parbox{2.5cm}{ \vskip 4pt Tangent sphere \vskip 3pt} & 1 quartuple (real) & $V.$ \\ \hline
\end{tabular}
\caption{the nine types of inequivalent rotational $R$-separable webs}
\end{table}

The remaining coordinates systems of Table 1 are equivalent to one of the coordinates listed above (correcting a typographical error in \cite{Moon}, where Cap cyclide coordinates are said to be equivalent to Bicyclide coordinates), as it is described in Table 3.
\begin{table}[h]
\begin{tabular}{|l|c|l|}
\hline
  Web & equivalent to & transformation \\ [9pt]
 \hline
Cap cyclide &  Flat-ring cyclide & cont. inversion + trans.  \\[9pt]  \hline
Prolate Spheroidal & Inverse Prolate Spheroidal & discrete inversion\\ [9pt]  \hline
Oblate Spheroidal & Inverse Oblate Spheroidal & discrete inversion\\[9pt] \hline
 Spherical & Bispherical & cont. inversion + trans. \\[9pt]  \hline
 Parabolical & Cardioid & discrete inversion \\[9pt] \hline
 Circular Cylindrical & Tangent sphere & discrete inversion\\[9pt]  \hline
\end{tabular}
\caption{Pairwise conformally equivalent webs}
\end{table}

\begin{rmk} \rm
The number of the types of rotationally $R$-separable coordinate systems agree with the results of \cite{Boyer}, where the subject is examined from the point of view of symmetry operators. The coefficients $A^{ij}$ of the second order part of the symmetry operators $S$ characterizing each type of $R$-separable rotationally symmetric coordinates, with respect to Cartesian coordinates, listed in Table 2. of Boyer, et al.
\cite{Boyer} when written as 
$$S=A^{ij}\partial_{i}\partial_{j}+ B^i\partial_i$$ correspond to the components of CKTs equivalent to those listed
in Table 1 for Bi-cyclide, Flat-ring cyclide, Disk cyclide and Toroidal coordinates, respectively.
\end{rmk}

Finally, we provide algebraic conditions on the parameters (\ref{par5}) in order to determine the type
of the corresponding web.
In order to obtain these conditions, we solve the equivalent problem of determining
the number and multiplicity of the linear factors of the corresponding binary quartic form $Q$ which can be done by
applying the classical algorithm (see for example \cite{Gurevich}) based on the sign and vanishing of relative invariants and covariants of $Q$.

Together with $I$ and $J$, the following invariant and covariants are used in the classification scheme: 
the discriminant of the form (a relative invariant which vanishes if and only if the quartic has a multiple root)
$$
\Delta= I^3-27J^2,
$$
the Hessian of the form (a covariant which vanishes if and only if the quartic has a quadruple root)
$$
H(X,Y)= (\partial^2_{XX}Q)\cdot(\partial^2_{YY}Q) - (\partial^2_{XY}Q)^2;
$$ 
the covariants
$$
L(X,Y)= I H(X,Y) - 6J Q(X,Y), 
$$
and
$$
M(X,Y)=12 H^2(X,Y) - I Q^2(X,Y). 
$$

We summarize the classification in Table 4.
\begin{table}[h]
\begin{tabular}[c]{|l|c|}
\hline
 Web &  Algebraic condition \\ [9pt]
 \hline
 Disk cyclide   & $\Delta <0$ \\[6pt]  
  \hline
 Bi-cyclide & $\Delta>0$ and $H(X,Y)<0$, and $M(X,Y)>0$  \\[6pt] 
 \hline
 Flat-ring cyclide&  $\Delta>0$ and ($H(X,Y)>0$ or $M(X,Y)>0$) \\ [6pt]  
 \hline
 Inverse prolate spheroidal & $\Delta=0$ and $L(X,Y)<0$\\ 
  [6pt]  
 \hline
 Inverse oblate spheroidal & $\Delta=0$ and $L(X,Y)>0$ \\[6pt] 
 \hline
 Toroidal & $L(X,Y)=0$ and $H(X,Y)>0$ \\ [6pt]  \hline
 Bispherical & $L(X,Y)=0$ and $H(X,Y)<0$ \\[6pt]  \hline
 Cardioid &  $I=J=0$ and $H(X,Y)\neq 0$ \\
[6pt] \hline
 Tangent sphere & $H(X,Y)=0$ \\[6pt]  \hline
\end{tabular}
\caption{Invariant classification of the webs}
\end{table}

The following simple example shows how the above method can be used.
\begin{exm} \rm
Let us consider the natural Hamiltonian scalar potential
\begin{equation}\label{exh}
H = \frac 12 (p_x^2+p_y^2+p_z^2)- \frac{4c^2}{(x^2+y^2+z^2-c^2)+4c^2z^2}.
\end{equation}
Although the potential clearly admits a rotational symmetry around the $z$-axis, it is straightforward
to check that the Hamilton-Jacobi equation $H=h$ does not admit separation of variables in any simply separable coordinate system of $\mathbb E^3$. This is because the compatibility condition (\ref{dKdV}) is never satisfied except for KTs of the form $\mathbf K= a\mathbf R_3\odot\mathbf R_3+b\mathbf g$, none of which are characteristic.
However, by imposing the conformal compatibility condition (\ref{Vcompat}) to the general rotationally symmetric CKT
(\ref{rotCKT}) we get, that for $E=0$, the conditions on the parameters of $\mathbf K$ are 
$$
M_{33}=\frac{1}{2c^2}H,\qquad L_{3} = 0,\qquad D_3 = 0,\qquad A_{33} = \frac {c^2}{2}H.
$$
The associated polynomial and binary quartic are
$$
\frac{z^4}{2c^2} + z^2 + \frac {c^2}{2}= \frac 12\left(\frac{z^2}c +c\right)^2, 
\quad Q(X,Y)=\frac{X^4}{2c^2} + X^2Y^2 + \frac {c^2}{2}Y^4,
$$
respectively. The polynomial has two double imaginary roots and - according to Table 2 - the associated web is the Toroidal
web. From the invariant point of view, the discriminant $\Delta$ and the covariant $L(X,Y)$ vanish and 
the Hessian of $Q$ is $H(X,Y)=12\frac{(X^2+Y^2\cdot{c^2})^2}{c^2}>0$.
Hence the Hamilton-Jacobi equation $H=0$, for the Hamiltonian (\ref{exh}) admits separation of variables in Toroidal
coordinates centered at the origin.
\end{exm}

\section{The remaining symmetric $R$-separable webs in $\mathbb{E}^{3}$}
In order to determine further CKTs which admit other types of symmetry we apply Proposition \ref{web-symm_2} to other types of CKVs: namely the translation along the $z$-axis $\mathbf{X}_3$, the dilatation $\mathbf{D}$ and the infinitesimal inversion along the $z$-axis $\mathbf I_3$. The three types of CKVs have been put in the above simple canonical forms by use of the isometry
group $SE(3)$.  

In all cases the infinitesimal invariance together with the TSN conditions are equivalent to the fact that
the infinitesimal symmetry is an eigenvector of the CKT (see Remark \ref{r_CVeigenv}).

The results obtained are given in the following proposition.
\begin{prop}
There exist no $R$-separable webs in $\mathbb{E}^{3}$ that possess translational, dilational or inversional symmetry which are not separable webs or can be obtained from one by the inversion $I$.
\end{prop}
\begin{proof}
We consider separately the three possible cases.\\
\textit{Translational symmetry}:

The equation (\ref{web-symm_3}) with $\mathbf V = \mathbf X_3$, $\mathbf K$ given by (\ref{CKT_compts_2}), and
$h = 0$, implies that the only non-zero components of the CKT are given by 
\begin{eqnarray}\label{formula57}
K_{11} &=& -A_{22}-A_{33}+1/2B_{21}z-1/2B_{31}y+C_{33}(y^2 + z^2)\nonumber\\
K_{22} &=& A_{22}-B_{21}z-1/2B_{31}y + C_{33}(y^2-2z^2)\nonumber\\
K_{33} &=& A_{33}+1/2B_{21}z+B_{31}y+C_{33}(z^2 - 2y^2)\nonumber\\
K_{23} &=& A_{23}-3/4B_{31}z+3/4B_{21}y+3C_{33}zy\nonumber\\
\end{eqnarray}
By comparison with (\ref{KT_compts}) we conclude that $\mathbf K$ is the trace free part of an ordinary KT and 
hence defines the simply separable webs which admit a translational symmetry along the z-axis \cite{CMP}. This result
agrees with that in \cite{Spencer_a} where a first principles proof is given that non-trivial $R$-separability of the Laplace and Helmholtz equations is never possible in a cylindrical coordinate system.\\
\textit{Dilatational symmetry}:\\
The equation (\ref{web-symm_3}) with $\mathbf V = \mathbf D$, $\mathbf K$ given by (\ref{CKT_compts_2}), and $h$ assumed constant, implies that $h$ has five possible integer values.   
The components of the CKT corresponding to each value of
$h$ are polynomials of the same degree. 
The imposition of the TSN condition yields an additional restriction only in the case when the components are second degree polynomials.  
All cases correspond either to
simply separable webs or can be mapped into one by the inversion $I$.  Thus we conclude that modulo equivalence with separable webs,
there are no $R$-separable webs which admit a dilatational symmetry.  \\
\textit{Inversional symmetry}:\\
The equation (\ref{web-symm_3}) with $\mathbf V = \mathbf I_3$, $\mathbf K$ given by (\ref{CKT_compts_2}), and $h$ some non-zero function implies that $\mathbf K = 0$.  If $h = 0$, it may be shown by use of the discrete inversion
that this case is equivalent to that of the translational symmetry considered above.  Thus, modulo equivalence, there are 
no $R$-symmetric webs admitting an inversional symmetry.
\end{proof}
We note that this proposition is in agreement with the
results of Table 2 of \cite{Boyer} where the only non-rotationally symmetric $R$-separable coordinate systems listed are the asymmetric cases not studied in this paper.

\section{Conclusions}
\setcounter{equation}{0}
We gave a classification of the rotationally symmetric $R$-separable webs for the Laplace equation on $\mathbb{E}^3$
in terms of the invariants of characteristic conformal Killing tensors under the action of the conformal group.  
Our method shows that there are exactly nine inequivalent types of webs, five of which are conformally equivalent to separable webs, in agreement with the results of
B\"{o}cher \cite{Bocher} and Boyer, Kalnins and Miller \cite{Boyer}. An invariant classification of the 
asymmetric webs based on the results of \cite{KeM} and \cite{Chanach} and following the approach of the present article will be the subject of a subsequent paper.

\section{Acknowledgments}
The authors wish to thank G. Rastelli and L. Degiovanni for helpful discussions on the background theory. MC and RGM
also wish to thank the the Department of Mathematics, Universit\`a di Torino for hospitality during which part of this
paper was written. The research was supported in part by a Natural Sciences and Engineering Research Council of
Canada Discovery Grant and by Senior Visiting Professorships of the Gruppo di Fisica Matematica dell'Italia (RGM)
and by a PRIN project of the Ministero dell'Universit`a e della Ricerca (CMC).



\end{document}